\documentclass[letter]{aa} 

\usepackage{graphicx}
\usepackage{txfonts}

\def\hi{\relax \ifmmode {\mbox H\,\texjtsc{i}}\else H\,{\scshape i}\fi}
\def\hii{\relax \ifmmode {\mbox H\,\textsc{ii}}\else H\,{\scshape ii}\fi}
\def\nii{\relax \ifmmode {\mbox N\,\textsc{ii}}\else N\,{\scshape ii}\fi}
\def\oi{\relax \ifmmode {\mbox O\,\textsc{i}}\else O\,{\scshape i}\fi}
\def\oii{\relax \ifmmode {\mbox O\,\textsc{ii}}\else O\,{\scshape ii}\fi}
\def\oiii{\relax \ifmmode {\mbox O\,\textsc{iii}}\else O\,{\scshape iii}\fi}
\def\sii{\relax \ifmmode {\mbox S\,\textsc{ii}}\else S\,{\scshape ii}\fi}

\begin{document}

\title{More fundamental than the fundamental metallicity relation:}
\subtitle{The effect of the stellar metallicity on the gas-phase mass-metallicity \\and gravitational potential-metallicity relations}
\titlerunning{The effect of the stellar metallicity on the MZR and $\Phi$MR}

\author{Laura S\'anchez-Menguiano\inst{1,2}
\and
Sebasti\'an F. S\'anchez\inst{3}
\and
Jorge S\'anchez Almeida\inst{4,5}
\and
Casiana Mu\~noz-Tu\~n\'on\inst{4,5}
          }
\authorrunning{L.~S\'anchez-Menguiano et al.}

\institute{Dpto. de F\'isica Te\'orica y del Cosmos, Facultad de Ciencias (Edificio Mecenas), Universidad de Granada, E-18071 Granada, Spain
\and
             Instituto Carlos I de F\'isica Te\'orica y computacional, Universidad de Granada, E-18071 Granada, Spain\\
              \email{lsanchezm@ugr.es}
\and
             Universidad Nacional Aut\'onoma de M\'exico, Instituto de Astronom\'ia, AP 106, Ensenada 22800, BC, M\'exico
\and
            Instituto de Astrof\'isica de Canarias, E-38205 La Laguna, Tenerife, Spain
\and
            Universidad de La Laguna, Dpto. Astrof\'isica, E-38206 La Laguna, Tenerife, Spain
             }

 \date{Received 30 October 2023 / Accepted 29 January 2024}

 
  \abstract
{One of the most fundamental scaling relations in galaxies is observed between metallicity and stellar mass -- the mass-metallicity relation (MZR) -- although recently a stronger dependence of the gas-phase metallicity with the galactic gravitational potential ($\Phi \rm ZR$) has been reported. Further dependences of metallicity on other galaxy properties have been revealed, with the star formation rate (SFR) being one of the most studied and debated secondary parameters in the relation (the so-called fundamental metallicity relation).}
{In this work we explore the dependence of the gas-phase metallicity residuals from the MZR and $\Phi \rm ZR$ on different galaxy properties in the search for the most fundamental scaling relation in galaxies.}
{We applied a random forest regressor algorithm on a sample of 3430 nearby star-forming galaxies from the SDSS-IV MaNGA survey. Using this technique, we explored the effect of 147 additional parameters on the global oxygen abundance residuals obtained after subtracting the MZR. Alternatively, we followed a similar approach with the metallicity residuals from the $\Phi \rm ZR$.}
{The stellar metallicity of the galaxy is revealed as the secondary parameter in both the MZR and the $\Phi \rm ZR$, ahead of the SFR. This parameter reduces the scatter in the relations $\sim 10-15\%$. We find the 3D relation between gravitational potential, gas metallicity, and stellar metallicity to be the most fundamental metallicity relation observed in galaxies.}
   {}


   \maketitle


\section{Introduction}


Scaling relations constitute a fundamental tool to improve our understanding on the formation and evolution of galaxies. A primary scaling relation links the gas-phase metallicity ($Z_g$) and the stellar mass \citep[MZR,][]{tremonti2004, kewley2008, wu2016, barreraballesteros2017, sanchez2019b, yates2020}. Its shape is characterised by a positive correlation between both parameters that flattens up at high masses. Confirmed up to redshift $z\sim10$  thanks to the advent of JWST \citep[presenting a similar shape but with a downward offset to lower $Z_g$ in high-z galaxies, e.g.][]{curti2023, nakajima2023}, the MZR has been proposed to result from the interplay of different processes including metal removal by outflows, dilution by metal-poor gas infall, or enrichment by previous star formation. A simple evolution by the so-called `downsizing', where most massive galaxies exhaust their gas reservoir faster, would also explain its origin \citep{maiolino2019}. 

Notable attention has been paid to the study of the scatter in the MZR, whose physical origin can unveil the relative importance of key processes in galaxy evolution. Thus, several secondary dependences in the MZR have been reported, such as that with the star formation rate (SFR) or specific SFR, gas fraction or gas mass, galaxy size, and stellar age \citep[e.g.][]{ellison2008, laralopez2010, lian2015, bothwell2016, sanchezalmeida2018, sanchezalmeida2019, curti2020, alvarezhurtado2022, scholte2023}, to name a few. Particularly relevant is the relation between the gas metallicity, stellar mass ($M_\star$), and SFR, called the fundamental metallicity relation \citep[FMR;][]{mannucci2010}, which shows how for low- and intermediate-mass galaxies ($\log[M_\star/M_\odot] < 10.5$) there exists an anti-correlation between the SFR and $Z_g$ at a given $M_\star$. The FMR is ascribed to the accretion of metal-poor cosmic gas fueling star formation \citep{mannucci2010, dave2012, sanchezalmeida2014}. However, there is a current debate on whether the SFR is truly needed to describe the relation between $M_\star$ and $Z_g$ \citep[e.g.][]{izotov2014, delosreyes2015, barreraballesteros2017, sanchez2017, duartepuertas2022}. 

Recent studies have also revealed a stronger dependence of $Z_g$ with the baryonic gravitational potential ($\Phi_{\rm baryon}$) than with $M_\star$ \citep[$\Phi \rm ZR$,][]{deugenio2018, sanchezmenguiano2024}. In particular, in \citet[][hereafter Paper I]{sanchezmenguiano2024}, we applied a random forest (RF) regressor on a sample of $\sim 3500$ nearby galaxies from the Mapping
Nearby Galaxies at Apache Point Observatory (MaNGA) survey and included, in the model, around 140 galaxy properties. Among the analysed parameters, we can find those that have been previously reported to strongly correlate with $Z_g$, such as $M_{\star}$, $\Phi_{\rm baryon}$, $\Sigma_{\star}$ and, $M_{\rm gas}$. We reported the $\Phi$ZR to be the tightest relation involving $Z_g$, and so the one with higher chances of being the primary one. 

In Paper I, we also show that a model including only $\Phi_{\rm baryon}$ or $M_{\star}$ as the input parameters ($\Phi$ZR and MZR, respectively) presented a root mean square error (RMSE) of the difference between the predictions (modelled $Z_g$) and the targets (measured $Z_g$) much higher than the complete model with all the input parameters. This result provides evidence that other parameters besides the gravitational potential (and stellar mass) play a significant role in shaping the global gas metallicity. Whether this parameter is the SFR or any other has yet to be found. Therefore, in order to investigate which is the most significant secondary dependence in the $\Phi$ZR and MZR, we analyse in this study the residuals of the $Z_g$ once the dependence with $\Phi_{\rm baryon}$ (and, alternatively, $M_{\star}$) has been modelled and subtracted out. Based on the use of the RF regressor and an extensive set of galaxy properties, we determine which of them present the tightest correlation with these metallicity residuals. As argued, this allows us to unveil the relative importance of key processes in galaxy evolution.

This Letter is organised as follows. Section~\ref{sec:manga} briefly presents the MaNGA data, the subset of galaxies comprising our study sample, and the physical parameters included in the model. A brief overview of the RF algorithm is given in Section~\ref{sec:RF}. We describe the outcome of the RF in Section~\ref{sec:results}. Finally, the results are discussed in Section~\ref{sec:dis}, and Section~\ref{sec:concl} compiles and summarises the main conclusions of the work. Supplementary material includes Appendices \ref{sec:appendix1} and \ref{sec:appendix2}. Appendix~\ref{sec:appendix1} reproduces the analysis using alternative calibrations to estimate $Z_g$. Appendix~\ref{sec:appendix2} describes the analysis for the $\Phi$ZR based on an alternative tracer of the total gravitational potential (including the effect of the dark matter halo).


\section{Sample and analysed data}\label{sec:manga}
This study is based on the data collected by the MaNGA project \citep[][]{bundy2015}, which is part of the fourth generation Sloan Digital Sky Survey (SDSS-IV). A very brief overview of the data is given in Paper I, and further details on the MaNGA mother sample, survey design, observational strategy, and data reduction are provided in the literature \citep{yan2016, wake2017, law2015, law2016, law2021}.

In order to investigate the existence of any secondary dependences in the $\Phi$ZR and MZR, we explored a list of 148 galaxy parameters extracted from the {\tt pyPipe3D} Value Added Catalogue \citep[VAC,][]{sanchez2022}, which is publicly accesible through the SDSS-IV VAC website\footnote{\url{https://www.sdss.org/dr17/data_access/value-added-catalogs/}}. We refer the reader to Paper I for further details on the selected physical properties, which are listed in appendix~A of that article. This list includes parameters such as $M_{\star}$, $\Phi_{\rm baryon}$, $\Sigma_{\star}$, $M_{\rm gas}$, morphological type, colours, SFR, stellar age and metallicity, or dust extinction. We used the oxygen abundance (O/H) measured at one disc effective radius ($R_e$) as a proxy for $Z_g$. This estimation is based on the O3N2 index and the empirical calibration proposed by \citet[][]{marino2013}. The results described in this Letter (Sec.~\ref{sec:results}) are confirmed applying alternative calibrators (see Appendix~\ref{sec:appendix1}). 

The analysed galaxy sample has been drawn from the 10\,010 galaxies comprising the MaNGA mother sample. As described in Paper I, for selecting the galaxies, we have adopted two simple criteria: galaxies have to meet the required quality standards of the analysis pipeline (i.e. QCFLAG field equal to zero in the {\tt pyPipe3D} VAC table, see section 4.5 of \citealt{sanchez2022} for details), and they must contain values for all of the analysed parameters. The second condition is quite restrictive. It excludes early-type systems and galaxies in general with minimal or negligible levels of ionised gas, making it impossible to derive gas-related attributes such as oxygen abundance or dust extinction. The resultant sample consists of 3\,430 galaxies, a sufficiently large number to guarantee the statistical robustness of the results. However, we note that the sample is biased towards large star-forming galaxies (mostly Sa to Sc galaxies).


\section{The RF regressor}\label{sec:RF}
The use of RF regressors \citep[][]{breiman2001} in the study of scaling relations has grown over the last years, offering a very compelling tool to unveil correlations between galaxy properties and identify the most significant dependencies \citep[e.g.][among others]{sanchezmenguiano2019, bluck2020, moster2021, piotrowska2022, baker2023}. For that, this algorithm employs an ensemble of decision trees to identify the input features (galaxy properties in this context) that carry the most complete information on the target feature (the galaxy parameter of interest, the residual gas metallicity in our case). Subsequently, it constructs a predictive model by establishing a set of conditions based on the values of the input features, gathering information on the relative importance of each property in the model in the process.

For this study we used a RF regressor to investigate the existence of any secondary dependences in the $\Phi$ZR and MZR. For that, we analysed the residuals of $Z_g$ once the dependence on $\Phi_{\rm baryon}$ (and, alternatively, $M_{\star}$) was modelled and subtracted. The RF algorithm was implemented in the {\tt scikit-learn} package for Python \citep{pedregosa2011}. While in Paper I we provide a brief overview of the basic steps involved and of the selected values for the model parameters, we refer the reader to the {\tt scikit-learn} User Guide documentation for comprehensive details on the complete algorithmic implementation\footnote{\url{https://scikit-learn.org/stable/modules/ensemble.html\#forest}}.


\section{Results}\label{sec:results}
\begin{figure*}
\centering
\resizebox{\hsize}{!}{\includegraphics{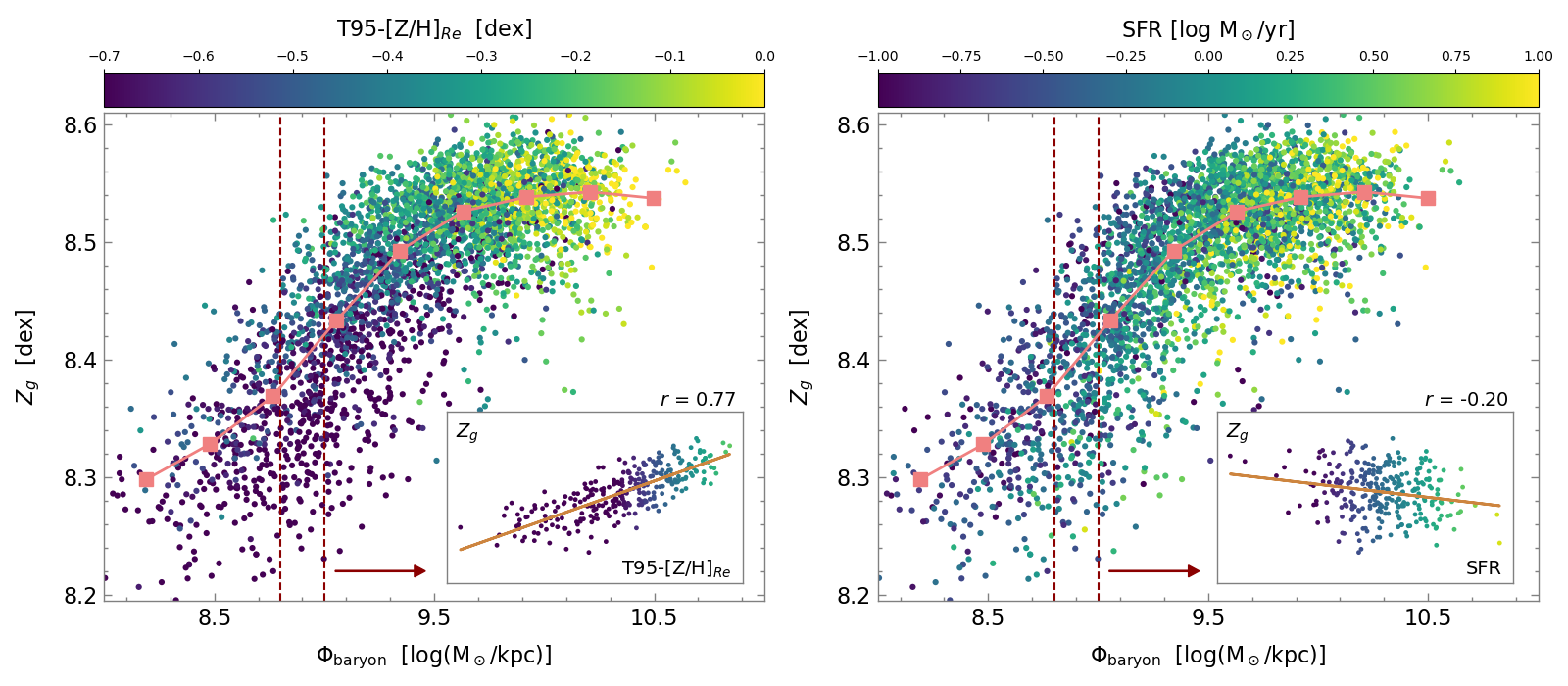}}
\caption{Global gas metallicity as a function of the galactic gravitational potential ($\Phi$ZR) colour-coded with the stellar metallicity measured at 1 $R_e$ and at the look-back time T95 (T95-[Z/H]$_{Re}$, {\it left panel}), and with the SFR ({\it right panel}). Salmon squares represent the median $Z_g$ values in ten bins. The averaged-binned values were fitted with a spline function (solid salmon lines).  The insets show a scatter plot $Z_g$ vs T95-[Z/H]$_{Re}$ ({\it left}) and $Z_g$ vs SFR ({\it right}) for the galaxies in the bin $8.8 \leq \Phi_{\rm baryon} \leq 9.0$, which is marked with vertical dashed lines in the main figure. The solid brown lines represent an ODR fit to the scatter points. The Pearson correlation coefficient $r$ is also displayed.}
\label{fig1}
\end{figure*}

The relation between $Z_g$ and $\Phi_{\rm baryon}$, the so-called $\Phi$ZR studied in Paper I, is represented in the left panel of Figure~\ref{fig1}. We can see that $Z_g$ increases when increasing $\Phi_{\rm baryon}$, with a flattening at the high end. In order to study secondary dependences on such a relation, we need to model the primary dependence of $Z_g$ on $\Phi_{\rm baryon}$. For that, the median values in ten bins of $\sim0.1$ dex width (salmon squares) were fitted with a spline function (solid salmon line). We see a reduction of $42\%$ in the dispersion with respect to that of the original relation. The parameter quantifying the reduction (denoted as $\sigma_{\rm red}$) was derived by subtracting the standard deviation of the residuals from the fit (difference between the measured $Z_g$ and the fit) to the standard deviation of the measured $Z_g$ values. This was then multiplied by 100 and divided by the standard deviation of the measured $Z_g$ values to obtain a percentage 

\begin{equation}\label{eq1}
\sigma_{\rm red} \, [\%] = 100 \cdot \frac{\sigma \,(Z_g) - \sigma \,(Z_g - Z_{g,fit})}{\sigma \,(Z_g)},
\end{equation}
where $Z_g$ is the measured global gas metallicity and $Z_{g,fit}$ is the modelled metallicity.
 
\begin{figure}
\centering
\resizebox{\hsize}{!}{\includegraphics{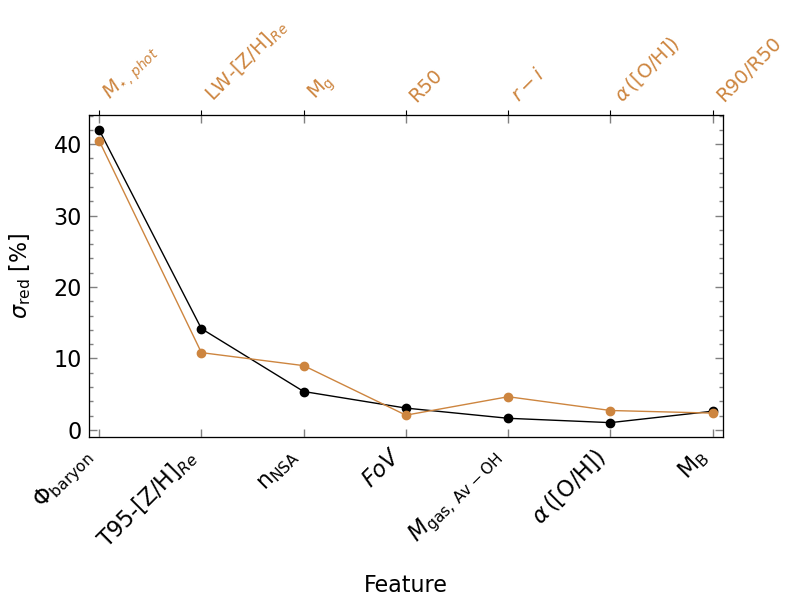}}
\caption{Decrease in the dispersion (in percent) of the original or residual global gas metallicity once the subsequent relation with different galaxy properties (x-axis) was modelled (with a spline function on binned values) and removed. This exercise is accumulative, so the modelled relation was subtracted to the residuals from the previous one. The secondary x-axis (top) shows the case when we forced the stellar mass to be the first parameter to model (orange line).}
\label{fig2}
\end{figure}

We subsequently ran the RF using the residual $Z_g$ ($\Delta Z_g$) as the targets and removing $\Phi_{\rm baryon}$ from the input parameters. The algorithm generated a model with T95-[Z/H]$_{Re}$ as the most relevant feature, which corresponds to the stellar metallicity at 1 $R_e$ measured at time T95 (the look-back time at which the galaxy formed 95\% of its mass, typically within the last 1 Gyr). T95-[Z/H]$_{Re}$ presents an importance value of 0.19\footnote{This parameter is a measure of how effective a feature is at reducing variance when splitting the variables along the decision trees. The higher the value is, the more important the feature.} and is followed in the model by the Sersic index (importance value of 0.06). The remaining parameters all have importance values below 0.03. This analysis shows that the $\Phi$ZR seems to present a  secondary dependence on the stellar metallicity. To quantify this dependence and its significance, we fitted the relation between $\Delta Z_g$ and T95-[Z/H]$_{Re}$ using again a spline function, and derived the decrease in the dispersion of $\Delta Z_g$ following eq.~(\ref{eq1}). We obtain a value of $14\%$. To explore further dependences of $Z_g$, we repeated this exercise fitting the residuals from the previous relation and running a RF with the remaining parameters. Figure~\ref{fig2} (black line, bottom x-axis) represents the decrease in the dispersion of $Z_g$ as we modelled its dependence with different galaxy parameters revealed by the RF. We can see that after fitting the dependence on $\Phi_{\rm baryon}$ ($42\%$) and the stellar metallicity ($14\%$), the fit with the remaining galaxy properties yields a decrease in the dispersion that falls below $5\%$ in all cases. This result suggests that the global gas metallicity can be unambiguously characterised by two parameters: the galactic gravitational potential and the stellar metallicity (in particular, T95-[Z/H]$_{Re}$). In the left panel of Figure~\ref{fig1}, the $\Phi$ZR is colour-coded by T95-[Z/H]$_{Re}$. We clearly see how, at a given value of the gravitational potential, $Z_g$ increases with increasing T95-[Z/H]$_{Re}$ (the inset shows this correlation for the particular bin $8.8 \leq \Phi_{\rm baryon} \leq 9.0$), highlighting the existence of this 3D relation between these parameters. The rest of galaxy properties do not seem to have a significant effect on $Z_g$. 

\begin{figure*}
\centering
\resizebox{\hsize}{!}{\includegraphics{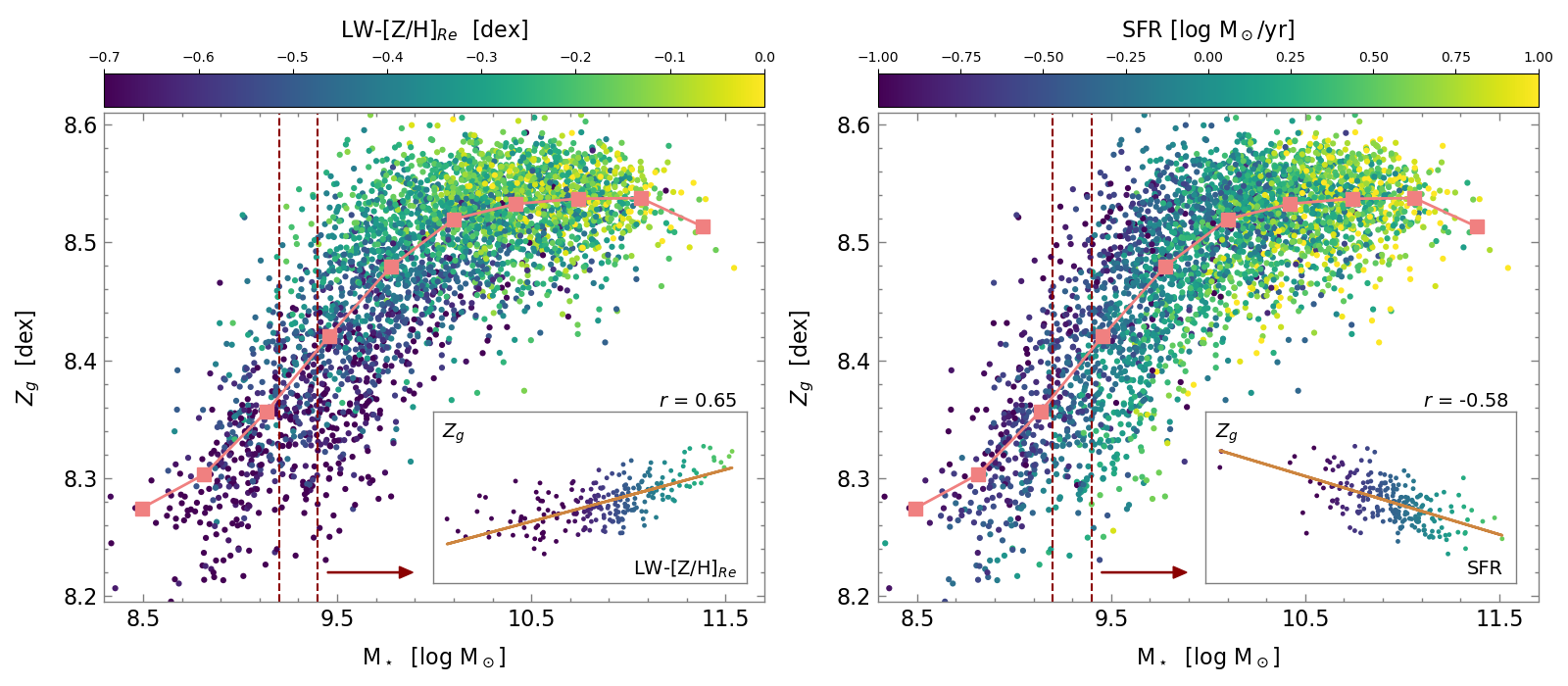}}
\caption{Global gas metallicity as a function of the stellar mass (MZR) colour-coded with the LW-stellar metallicity measured at 1 $R_e$ (LW-[Z/H]$_{Re}$, {\it left panel}), and with the SFR ({\it right panel}). Salmon squares represent the median $Z_g$ values in ten mass bins. The averaged-binned values were fitted with a spline function (solid salmon lines). The insets show a scatter plot $Z_g$ vs LW-[Z/H]$_{Re}$ ({\it left}) and $Z_g$ vs SFR ({\it right}) for the galaxies in the mass bin $9.2 \leq \log({\rm M_{\star}}) \leq 9.4$, which is marked with vertical dashed lines in the main figure. The  solid brown lines represent an ODR fit to the scatter points. The Pearson correlation coefficient $r$ is also displayed.}
\label{fig3}
 \end{figure*}

Alternatively, we repeated this exercise based on the residuals from subtracting the dependence of $Z_g$ with $M_{\star}$. This will allow us to look for secondary dependences in the MZR. The results are shown in Figure~\ref{fig2} (orange line, top x-axis). We can see that the MZR reduces the dispersion in the $Z_g$ values to a lower extent ($40\%$ as opposed to $42\%$, see Paper I). In this case, the highest effect on the relation comes from the luminosity-weighted stellar metallicity measured at 1 $R_e$ (LW-[Z/H]$_{Re}$), which is another indicator of the global stellar metallicity ($11\%$ decrease in the dispersion). The following parameter in the ranking is the absolute magnitude in the $g$ band ($9\%$), and the remaining ones present values below $3\%$. The left panel of Figure~\ref{fig3} represents the MZR (i.e. gas metallicity as a function of the stellar mass) colour-coded by LW-[Z/H]$_{Re}$. We clearly see how, given a stellar mass, $Z_g$ increases with increasing stellar metallicity (the inset shows this correlation for the particular bin $9.2 \leq \log({\rm M_{\star}}) \leq 9.4$), revealing the existing relation between these parameters.

In order to compare the analysed relations with the FMR, the right panels of Figure~\ref{fig1} and Figure~\ref{fig3} show the dependence of both $\Phi$ZR and MZR on the SFR. At a given value of $\Phi_{\rm baryon}$ and $M_{\star}$, the colour-coding and the insets reflect a decrease in gas metallicity when SFR increases. We  thus checked how much of a reduction in the dispersion would result by fitting the SFR as the secondary parameter in both relations. Whereas the stellar metallicity produces a reduction of $\sim11-14\%$, this amount falls below $3\%$ when considering the SFR. Since the shape of the dependence of the MZR with SFR depends on the considered stellar-mass range \citep{mannucci2010, sanchezalmeida2019}, we performed several tests by dividing the sample into different mass ranges and determining the reduction in the dispersion on these subsamples. Table~\ref{tab2} summarises the results. We can see that the dependence with the SFR is higher in the MZR, and quite significant for masses below $\sim 10 \, \log \, M_\odot$. However, the dependence with LW-[Z/H]$_{Re}$ is stronger in all mass ranges. In the case of the $\Phi$ZR, the effect of the SFR on the relation is negligible in all ranges of gravitational potential, whereas the dependence with T95-[Z/H]$_{Re}$ is clear.

\begin{table*}
\caption{Decrease in the dispersion (in percent) of the $\Phi$ZR and MZR when considering a secondary relation with the SFR in comparison with an indicator of the stellar metallicity for different ranges of mass (for MZR) or gravitational potential (for $\Phi$ZR). See main text for details.}             
\label{tab2}      
\centering          
\begin{tabular}{r c c c c c c c}     
 &   &   &   &  $\Phi$ZR &   &  &  \\[0cm]
\hline\hline\\[-0.25cm]
 & All & $\left[8.2, 8.4\right]$ & $\left[8.6, 8.8\right]$ & $\left[9.0, 9.2\right]$ & $\left[9.4, 9.6\right]$ & $\left[9.8, 10.0\right]$ & $\left[10.2, 10.4\right]$ \\[0.1cm]
\hline \\[-0.25cm]        
   SFR, $\sigma_{\rm red}$ [\%] & $0.3$ & $4.1$ & $3.4$ & $1.0$ & $3.6$ & $4.0$ & $2.7$\\[0.1cm]
   T95-[Z/H]$_{Re}$, $\sigma_{\rm red}$ [\%] & $14.2$ & $22.7$ & $26.4$ & $30.4$ & $19.6$ & $2.4$ & $0.1$\\[0.1cm]
\hline\\[0.1cm]
 &   &   &   &  MZR &   &   &   \\[0cm]
\hline\hline\\[-0.25cm]
 & All & $\left[8.7, 8.9\right]$ & $\left[9.1, 9.3\right]$ & $\left[9.5, 9.7\right]$ & $\left[9.9, 10.1\right]$ & $\left[10.3, 10.5\right]$ & $\left[10.7, 10.9\right]$ \\[0.1cm]
\hline \\[-0.25cm]        
   SFR, $\sigma_{\rm red}$ [\%] & $2.3$ & $11.9$ & $19.7$ & $17.0$ & $4.8$ & $2.8$ & $1.7$ \\[0.1cm]
   LW-[Z/H]$_{Re}$, $\sigma_{\rm red}$ [\%] & $10.7$ & $9.5$ & $21.5$ & $31.6$ & $19.5$ & $6.6$ & $1.6$ \\[0.1cm]
\hline\\[0.1cm]
\end{tabular}
\end{table*}

\section{Discussion}\label{sec:dis}
The scatter in the MZR has been observed to correlate with several galaxy properties. The anti-correlation between the gas metallicity and SFR at a fixed stellar mass was reported in various studies, which defended the existence of a thin surface in 3D space defined by $M_{\star}$, $Z_g$, and SFR where star-forming galaxies are confined \citep{mannucci2010, laralopez2010, curti2020}. Its initially revealed lack of evolution with redshift\footnote{Although no evidence of redshift evolution has been observed up to $z \sim 3$ \citep[e.g.][]{sanders2021}, recent studies based on JWST data indicate a clear deviation from the local FMR especially for $z>6$ \citep[e.g.][]{curti2023b, curti2023}.} earned it the name fundamental metallicity relation, making it a key ingredient for chemical and galaxy evolution studies. Some works, nonetheless, argue that the secondary dependence on the SFR does not truly improve the relationship between $Z_g$ and $M_{\star}$, with the reduction in the scatter being insignificant \citep[e.g.][]{hughes2013, sanchez2019b}. Recent studies also suggest that, if it exists, the reported correlation between the MZR scatter and the SFR could be a by-product of a more primordial relation with the gas mass and/or fraction \citep{bothwell2013, bothwell2016, bothwell2016b, chen2022, scholte2023}, due to the well-known link between gas content and star formation activity.  

Regarding the $\Phi$ZR, no previous works have explored how the scatter might depend on different galaxy properties. In this study, we investigate for the first time this matter, as well as try to provide a definite answer as to the reported correlations of the MZR residuals. Using a RF regressor algorithm, we find that the MZR and the $\Phi$ZR residuals are best predicted by two estimators of the global stellar metallicity: LW-[Z/H]$_{Re}$ (for the MZR) and T95-$ \rm [Z/H]_{Re}$ (for the $\Phi$ZR). These parameters provide a reduction in the scatter of the primary relations of around $11\%$ and $14\%$, respectively. We note that the reduction in the scatter of the MZR provided by T95-$ \rm [Z/H]_{Re}$, LW-[Z/H]$_{Re}$, or even MW-[Z/H]$_{Re}$ are very similar, with differences below $0.5\%$. These differences, although not significant, make the RF prioritise one of these parameters as opposed to the others when developing the model. In any case, it is clear that the stellar metallicity, however estimated, is the secondary parameter with the highest effect on MZR and $\Phi$ZR. In contrast, the relative importance of parameters such as SFR or $M_{\rm gas}$\footnote{We note that $M_{\rm gas}$ was indirectly estimated using the dust extinction as a tracer via the dust-to-gas relation, whereas the SFR was determined from dust-corrected H$\alpha$ luminosity. A more direct or alternative measurement of the parameters might yield different results.} in the model is very low, with the reduction in the scatter for the case of the SFR being barely $0.5-2\%$. When we restricted the analysis to specific stellar mass ranges, the reduction in the scatter was higher for the stellar metallicity  in all cases. Nevertheless, we see an improvement in the reduction of the MZR scatter with the SFR up to $20\%$ for $\log(M_{\star})\sim 9 \, M_\odot$, in agreement with the original studies of the FMR that find a stronger anti-correlation between $Z_g$ and SFR for low-mass galaxies \citep{sanchezalmeida2019}. This is not the case for the $\Phi$ZR, where this percentage is always below $5\%$. We ascribe this difference to a size effect: while the size of the galaxies is accounted for in the gravitational potential (estimated as $M_{\star}/R_e$), it is not in the stellar mass. Therefore, at a fixed stellar mass, a more compact galaxy would present a higher gas density, and thus a higher SFR \citep{campsfarina2023}. This effect would be stronger at lower masses because the dynamical range of $R_e$ is larger. At a fixed $\Phi_{\rm baryon}$, since the size effect is accounted for, the dependence with the SFR disappears. 

Our results hold independently of the adopted $Z_g$ indicator. This is shown in Appendix~\ref{sec:appendix1}, where we reproduce the analysis with 15 alternative estimators, revealing that for most cases an indicator of the global stellar metallicity is the galactic property with the highest effect on the MZR and $\Phi$ZR residuals. In the remaining cases, we find a stronger dependence on the $Z_g$ gradient that we ascribe to the uncertainty in the determination of $R_e$, leading to a wrong estimation of the characteristic $Z_g$. 

The secondary dependence of the baryonic $\Phi$ZR on the stellar metallicity that was found is also confirmed when exploring the total $\Phi$ZR using $M_\star / R_e^{\,0.6}$ as a tracer of the total gravitational potential (the exponent $\alpha \simeq 0.6$ accounts for the inclusion of the dark matter component; see Paper I). In Appendix~\ref{sec:appendix2} we look at the total $\Phi$ZR residuals, removing the dependence of $Z_g$ on $M_\star / R_e^{\,0.6}$ instead of $\Phi_{\rm baryon}$, and we obtain very similar results to the latter, with the stellar metallicity being the parameter with the strongest effect on $\Delta Z_g$.

Previous studies attribute the dependence of the MZR scatter found on the SFR to the significant effect of recent gas accretion on the present-day chemical distribution of galaxies. Our results suggest otherwise. The strongest role played by the stellar metallicity in the RF models for predicting the MZR and $\Phi$ZR residuals provides evidence that the present-day gas metallicity is mostly driven by the overall chemical enrichment history (ChEH), and not the most recent episodes (of which the present SFR is indicative). In this sense, the fact that T95-$ \rm [Z/H]_{Re}$ or LW-[Z/H]$_{Re}$ are the estimators of the global stellar metallicity that best predict $\Delta Z_g$ reinforces this idea. The mean LW-[Z/H] enhances the contribution of the youngest stars compared with the mean MW-[Z/H], which makes it more sensitive to the ChEH because the content of old stars is similar in all galaxies (in the analysed mass range). In any case, as discussed, the differences obtained by considering different estimators of stellar metallicity are small; what is relevant is the role played by stellar metallicity over other galaxy properties on both relations.

\section{Summary and conclusions}\label{sec:concl}
We have analysed the existence of secondary dependences in the scaling relations between stellar mass and gas metallicity (MZR) and between gravitational potential and gas metallicity ($\Phi$ZR). Using a RF regressor, and based on a sample of 3430 star-forming MaNGA galaxies, we looked for the galactic property that best predicts the MZR and $\Phi$ZR residuals. We paid special attention to the role of the SFR in the model, which is claimed to be a primordial one according to the {fundamental metallicity relation}. 

We conclude that the global stellar metallicity is the galaxy property with the strongest effect on the MZR and $\Phi$ZR residuals. This is true independently of the different adopted indicators of the gas metallicity (Appendix~\ref{sec:appendix1}) and considering both the baryonic and the total $\Phi$ZR (Appendix~\ref{sec:appendix2}). This parameter reduces the scatter in the MZ and $\Phi$Z relations by $10-15\%$, whereas the SFR only reduces the scatter by $0.5-2\%$. For particular mass ranges (specially for $\log(M_{\star})\sim 9 \, M_\odot$), the reduction obtained with the SFR can improve up to $20\%$ for the MZR, but it is always below the decrease reached with the stellar metallicity. These results suggest that the present-day gas metallicity distribution is mostly affected by the overall chemical enrichment history of the galaxy, rather than recent events (of which the present SFR is representative) driven by gas accretion, as previously claimed in the literature.

\begin{acknowledgements}
LSM acknowledges support from Juan de la Cierva fellowship (IJC2019-041527-I) and from project PID2020-114414GB-100, financed by MCIN/AEI/10.13039/501100011033. S.F.S. thanks the PAPIIT-DGAPA AG100622 project. This research was partly funded by the Spanish Ministry of Science and Innovation, project PID2022-136598NB-C31 (ESTALLIDOS). \\[0.1cm]
This project makes use of the MaNGA-Pipe3D dataproducts. We thank the IA-UNAM MaNGA team for creating this catalogue, and the Conacyt Project CB-285080 for supporting them. 
\end{acknowledgements}

\bibliographystyle{aa} 
\bibliography{bibliography} 

\begin{thebibliography}{59}
\expandafter\ifx\csname natexlab\endcsname\relax\def\natexlab#1{#1}\fi

\bibitem[{{Alvarez-Hurtado} {et~al.}(2022){Alvarez-Hurtado},
  {Barrera-Ballesteros}, {S{\'a}nchez}, {Colombo}, {L{\'o}pez-S{\'a}nchez}, \&
  {Aquino-Ort{\'\i}z}}]{alvarezhurtado2022}
{Alvarez-Hurtado}, P., {Barrera-Ballesteros}, J.~K., {S{\'a}nchez}, S.~F.,
  {et~al.} 2022, \apj, 929, 47

\bibitem[{{Baker} {et~al.}(2023){Baker}, {Maiolino}, {Belfiore}, {Bluck},
  {Curti}, {Wylezalek}, {Bertemes}, {Bothwell}, {Lin}, {Thorp}, \&
  {Pan}}]{baker2023}
{Baker}, W.~M., {Maiolino}, R., {Belfiore}, F., {et~al.} 2023, \mnras, 518,
  4767

\bibitem[{{Barrera-Ballesteros} {et~al.}(2017){Barrera-Ballesteros},
  {S{\'a}nchez}, {Heckman}, {Blanc}, \& {The MaNGA
  Team}}]{barreraballesteros2017}
{Barrera-Ballesteros}, J.~K., {S{\'a}nchez}, S.~F., {Heckman}, T., {Blanc},
  G.~A., \& {The MaNGA Team}. 2017, \apj, 844, 80

\bibitem[{{Bluck} {et~al.}(2020){Bluck}, {Maiolino}, {S{\'a}nchez}, {Ellison},
  {Thorp}, {Piotrowska}, {Teimoorinia}, \& {Bundy}}]{bluck2020}
{Bluck}, A. F.~L., {Maiolino}, R., {S{\'a}nchez}, S.~F., {et~al.} 2020, \mnras,
  492, 96

\bibitem[{{Bothwell} {et~al.}(2016{\natexlab{a}}){Bothwell}, {Maiolino},
  {Cicone}, {Peng}, \& {Wagg}}]{bothwell2016b}
{Bothwell}, M.~S., {Maiolino}, R., {Cicone}, C., {Peng}, Y., \& {Wagg}, J.
  2016{\natexlab{a}}, \aap, 595, A48

\bibitem[{{Bothwell} {et~al.}(2013){Bothwell}, {Maiolino}, {Kennicutt},
  {Cresci}, {Mannucci}, {Marconi}, \& {Cicone}}]{bothwell2013}
{Bothwell}, M.~S., {Maiolino}, R., {Kennicutt}, R., {et~al.} 2013, \mnras, 433,
  1425

\bibitem[{{Bothwell} {et~al.}(2016{\natexlab{b}}){Bothwell}, {Maiolino},
  {Peng}, {Cicone}, {Griffith}, \& {Wagg}}]{bothwell2016}
{Bothwell}, M.~S., {Maiolino}, R., {Peng}, Y., {et~al.} 2016{\natexlab{b}},
  \mnras, 455, 1156

\bibitem[{Breiman(2001)}]{breiman2001}
Breiman, L. 2001, Machine Learning, 45, 5

\bibitem[{{Bundy} {et~al.}(2015){Bundy}, {Bershady}, {Law}, {Yan}, {Drory},
  {MacDonald}, {Wake}, {Cherinka}, {S{\'a}nchez-Gallego}, {Weijmans}, {Thomas},
  {Tremonti}, {Masters}, {Coccato}, {Diamond-Stanic}, {Arag{\'o}n-Salamanca},
  {Avila-Reese}, {Badenes}, {Falc{\'o}n-Barroso}, {Belfiore}, {Bizyaev},
  {Blanc}, {Bland-Hawthorn}, {Blanton}, {Brownstein}, {Byler}, {Cappellari},
  {Conroy}, {Dutton}, {Emsellem}, {Etherington}, {Frinchaboy}, {Fu}, {Gunn},
  {Harding}, {Johnston}, {Kauffmann}, {Kinemuchi}, {Klaene}, {Knapen},
  {Leauthaud}, {Li}, {Lin}, {Maiolino}, {Malanushenko}, {Malanushenko}, {Mao},
  {Maraston}, {McDermid}, {Merrifield}, {Nichol}, {Oravetz}, {Pan}, {Parejko},
  {Sanchez}, {Schlegel}, {Simmons}, {Steele}, {Steinmetz}, {Thanjavur},
  {Thompson}, {Tinker}, {van den Bosch}, {Westfall}, {Wilkinson}, {Wright},
  {Xiao}, \& {Zhang}}]{bundy2015}
{Bundy}, K., {Bershady}, M.~A., {Law}, D.~R., {et~al.} 2015, \apj, 798, 7

\bibitem[{{Camps-Fari{\~n}a} {et~al.}(2023){Camps-Fari{\~n}a},
  {S{\'a}nchez-Bl{\'a}zquez}, {Roca-F{\`a}brega}, \&
  {S{\'a}nchez}}]{campsfarina2023}
{Camps-Fari{\~n}a}, A., {S{\'a}nchez-Bl{\'a}zquez}, P., {Roca-F{\`a}brega}, S.,
  \& {S{\'a}nchez}, S.~F. 2023, \aap, 678, A65

\bibitem[{{Chen} {et~al.}(2022){Chen}, {Wang}, \& {Kong}}]{chen2022}
{Chen}, X., {Wang}, J., \& {Kong}, X. 2022, \apj, 933, 39

\bibitem[{{Curti} {et~al.}(2017){Curti}, {Cresci}, {Mannucci}, {Marconi},
  {Maiolino}, \& {Esposito}}]{curti2017}
{Curti}, M., {Cresci}, G., {Mannucci}, F., {et~al.} 2017, \mnras, 465, 1384

\bibitem[{{Curti} {et~al.}(2023{\natexlab{a}}){Curti}, {D'Eugenio}, {Carniani},
  {Maiolino}, {Sandles}, {Witstok}, {Baker}, {Bennett}, {Piotrowska},
  {Tacchella}, {Charlot}, {Nakajima}, {Maheson}, {Mannucci}, {Amiri},
  {Arribas}, {Belfiore}, {Bonaventura}, {Bunker}, {Chevallard}, {Cresci},
  {Curtis-Lake}, {Hayden-Pawson}, {Jones}, {Kumari}, {Laseter}, {Looser},
  {Marconi}, {Maseda}, {Scholtz}, {Smit}, {{\"U}bler}, \&
  {Wallace}}]{curti2023b}
{Curti}, M., {D'Eugenio}, F., {Carniani}, S., {et~al.} 2023{\natexlab{a}},
  \mnras, 518, 425

\bibitem[{{Curti} {et~al.}(2023{\natexlab{b}}){Curti}, {Maiolino},
  {Curtis-Lake}, {Chevallard}, {Carniani}, {D'Eugenio}, {Looser}, {Scholtz},
  {Charlot}, {Cameron}, {{\"U}bler}, {Witstok}, {Boyett}, {Laseter}, {Sandles},
  {Arribas}, {Bunker}, {Giardino}, {Maseda}, {Rawle}, {Rodr{\'\i}guez Del
  Pino}, {Smit}, {Willott}, {Eisenstein}, {Hausen}, {Johnson}, {Rieke},
  {Robertson}, {Tacchella}, {Williams}, {Willmer}, {Baker}, {Bhatawdekar},
  {Egami}, {Helton}, {Ji}, {Kumari}, {Perna}, {Shivaei}, \& {Sun}}]{curti2023}
{Curti}, M., {Maiolino}, R., {Curtis-Lake}, E., {et~al.} 2023{\natexlab{b}},
  arXiv e-prints, arXiv:2304.08516

\bibitem[{{Curti} {et~al.}(2020){Curti}, {Mannucci}, {Cresci}, \&
  {Maiolino}}]{curti2020}
{Curti}, M., {Mannucci}, F., {Cresci}, G., \& {Maiolino}, R. 2020, \mnras, 491,
  944

\bibitem[{{Dav{\'e}} {et~al.}(2012){Dav{\'e}}, {Finlator}, \&
  {Oppenheimer}}]{dave2012}
{Dav{\'e}}, R., {Finlator}, K., \& {Oppenheimer}, B.~D. 2012, \mnras, 421, 98

\bibitem[{{de los Reyes} {et~al.}(2015){de los Reyes}, {Ly}, {Lee}, {Salim},
  {Peeples}, {Momcheva}, {Feddersen}, {Dale}, {Ouchi}, {Ono}, \&
  {Finn}}]{delosreyes2015}
{de los Reyes}, M.~A., {Ly}, C., {Lee}, J.~C., {et~al.} 2015, \aj, 149, 79

\bibitem[{{D'Eugenio} {et~al.}(2018){D'Eugenio}, {Colless}, {Groves}, {Bian},
  \& {Barone}}]{deugenio2018}
{D'Eugenio}, F., {Colless}, M., {Groves}, B., {Bian}, F., \& {Barone}, T.~M.
  2018, \mnras, 479, 1807

\bibitem[{{Duarte Puertas} {et~al.}(2022){Duarte Puertas}, {Vilchez},
  {Iglesias-P{\'a}ramo}, {Moll{\'a}}, {P{\'e}rez-Montero}, {Kehrig},
  {Pilyugin}, \& {Zinchenko}}]{duartepuertas2022}
{Duarte Puertas}, S., {Vilchez}, J.~M., {Iglesias-P{\'a}ramo}, J., {et~al.}
  2022, \aap, 666, A186

\bibitem[{{Ellison} {et~al.}(2008){Ellison}, {Patton}, {Simard}, \&
  {McConnachie}}]{ellison2008}
{Ellison}, S.~L., {Patton}, D.~R., {Simard}, L., \& {McConnachie}, A.~W. 2008,
  \apjl, 672, L107

\bibitem[{{Ho}(2019)}]{ho2019}
{Ho}, I.~T. 2019, \mnras, 485, 3569

\bibitem[{{Hughes} {et~al.}(2013){Hughes}, {Cortese}, {Boselli}, {Gavazzi}, \&
  {Davies}}]{hughes2013}
{Hughes}, T.~M., {Cortese}, L., {Boselli}, A., {Gavazzi}, G., \& {Davies},
  J.~I. 2013, \aap, 550, A115

\bibitem[{{Izotov} {et~al.}(2014){Izotov}, {Guseva}, {Fricke}, \&
  {Henkel}}]{izotov2014}
{Izotov}, Y.~I., {Guseva}, N.~G., {Fricke}, K.~J., \& {Henkel}, C. 2014, \aap,
  561, A33

\bibitem[{{Kewley} \& {Dopita}(2002)}]{kewley2002}
{Kewley}, L.~J. \& {Dopita}, M.~A. 2002, \apjs, 142, 35

\bibitem[{{Kewley} \& {Ellison}(2008)}]{kewley2008}
{Kewley}, L.~J. \& {Ellison}, S.~L. 2008, \apj, 681, 1183

\bibitem[{{Kewley} {et~al.}(2019){Kewley}, {Nicholls}, \&
  {Sutherland}}]{kewley2019}
{Kewley}, L.~J., {Nicholls}, D.~C., \& {Sutherland}, R.~S. 2019, \araa, 57, 511

\bibitem[{{Kobulnicky} \& {Kewley}(2004)}]{kobulnicky2004}
{Kobulnicky}, H.~A. \& {Kewley}, L.~J. 2004, \apj, 617, 240

\bibitem[{{Lara-L{\'o}pez} {et~al.}(2010){Lara-L{\'o}pez}, {Cepa},
  {Bongiovanni}, {P{\'e}rez Garc{\'{\i}}a}, {Ederoclite}, {Casta{\~n}eda},
  {Fern{\'a}ndez Lorenzo}, {Povi{\'c}}, \&
  {S{\'a}nchez-Portal}}]{laralopez2010}
{Lara-L{\'o}pez}, M.~A., {Cepa}, J., {Bongiovanni}, A., {et~al.} 2010, \aap,
  521, L53

\bibitem[{{Law} {et~al.}(2016){Law}, {Cherinka}, {Yan}, {Andrews}, {Bershady},
  {Bizyaev}, {Blanc}, {Blanton}, {Bolton}, {Brownstein}, {Bundy}, {Chen},
  {Drory}, {D'Souza}, {Fu}, {Jones}, {Kauffmann}, {MacDonald}, {Masters},
  {Newman}, {Parejko}, {S{\'a}nchez-Gallego}, {S{\'a}nchez}, {Schlegel},
  {Thomas}, {Wake}, {Weijmans}, {Westfall}, \& {Zhang}}]{law2016}
{Law}, D.~R., {Cherinka}, B., {Yan}, R., {et~al.} 2016, \aj, 152, 83

\bibitem[{{Law} {et~al.}(2021){Law}, {Westfall}, {Bershady}, {Cappellari},
  {Yan}, {Belfiore}, {Bizyaev}, {Brownstein}, {Chen}, {Cherinka}, {Drory},
  {Lazarz}, \& {Shetty}}]{law2021}
{Law}, D.~R., {Westfall}, K.~B., {Bershady}, M.~A., {et~al.} 2021, \aj, 161, 52

\bibitem[{{Law} {et~al.}(2015){Law}, {Yan}, {Bershady}, {Bundy}, {Cherinka},
  {Drory}, {MacDonald}, {S{\'a}nchez-Gallego}, {Wake}, {Weijmans}, {Blanton},
  {Klaene}, {Moran}, {Sanchez}, \& {Zhang}}]{law2015}
{Law}, D.~R., {Yan}, R., {Bershady}, M.~A., {et~al.} 2015, \aj, 150, 19

\bibitem[{{Lian} {et~al.}(2015){Lian}, {Li}, {Yan}, \& {Kong}}]{lian2015}
{Lian}, J.~H., {Li}, J.~R., {Yan}, W., \& {Kong}, X. 2015, \mnras, 446, 1449

\bibitem[{{L{\'o}pez-S{\'a}nchez} {et~al.}(2012){L{\'o}pez-S{\'a}nchez},
  {Dopita}, {Kewley}, {Zahid}, {Nicholls}, \&
  {Scharw{\"a}chter}}]{lopezsanchez2012}
{L{\'o}pez-S{\'a}nchez}, {\'A}.~R., {Dopita}, M.~A., {Kewley}, L.~J., {et~al.}
  2012, \mnras, 426, 2630

\bibitem[{{Maiolino} \& {Mannucci}(2019)}]{maiolino2019}
{Maiolino}, R. \& {Mannucci}, F. 2019, \aapr, 27, 3

\bibitem[{{Mannucci} {et~al.}(2010){Mannucci}, {Cresci}, {Maiolino}, {Marconi},
  \& {Gnerucci}}]{mannucci2010}
{Mannucci}, F., {Cresci}, G., {Maiolino}, R., {Marconi}, A., \& {Gnerucci}, A.
  2010, \mnras, 408, 2115

\bibitem[{{Marino} {et~al.}(2013){Marino}, {Rosales-Ortega}, {S{\'a}nchez},
  {Gil de Paz}, {V{\'{\i}}lchez}, {Miralles-Caballero}, {Kehrig},
  {P{\'e}rez-Montero}, {Stanishev}, {Iglesias-P{\'a}ramo}, {D{\'{\i}}az},
  {Castillo-Morales}, {Kennicutt}, {L{\'o}pez-S{\'a}nchez}, {Galbany},
  {Garc{\'{\i}}a-Benito}, {Mast}, {Mendez-Abreu}, {Monreal-Ibero}, {Husemann},
  {Walcher}, {Garc{\'{\i}}a-Lorenzo}, {Masegosa}, {Del Olmo Orozco},
  {Mour{\~a}o}, {Ziegler}, {Moll{\'a}}, {Papaderos},
  {S{\'a}nchez-Bl{\'a}zquez}, {Gonz{\'a}lez Delgado}, {Falc{\'o}n-Barroso},
  {Roth}, {van de Ven}, \& {Califa Team}}]{marino2013}
{Marino}, R.~A., {Rosales-Ortega}, F.~F., {S{\'a}nchez}, S.~F., {et~al.} 2013,
  \aap, 559, A114

\bibitem[{{Moster} {et~al.}(2021){Moster}, {Naab}, {Lindstr{\"o}m}, \&
  {O'Leary}}]{moster2021}
{Moster}, B.~P., {Naab}, T., {Lindstr{\"o}m}, M., \& {O'Leary}, J.~A. 2021,
  \mnras, 507, 2115

\bibitem[{{Nakajima} {et~al.}(2023){Nakajima}, {Ouchi}, {Isobe}, {Harikane},
  {Zhang}, {Ono}, {Umeda}, \& {Oguri}}]{nakajima2023}
{Nakajima}, K., {Ouchi}, M., {Isobe}, Y., {et~al.} 2023, arXiv e-prints,
  arXiv:2301.12825

\bibitem[{{Nestor Shachar} {et~al.}(2023){Nestor Shachar}, {Price},
  {F{\"o}rster Schreiber}, {Genzel}, {Shimizu}, {Tacconi}, {{\"U}bler},
  {Burkert}, {Davies}, {Dekel}, {Herrera-Camus}, {Lee}, {Liu}, {Lutz}, {Naab},
  {Neri}, {Renzini}, {Saglia}, {Schuster}, {Sternberg}, {Wisnioski}, \&
  {Wuyts}}]{nestorshachar2023}
{Nestor Shachar}, A., {Price}, S.~H., {F{\"o}rster Schreiber}, N.~M., {et~al.}
  2023, \apj, 944, 78

\bibitem[{Pedregosa {et~al.}(2011)Pedregosa, Varoquaux, Gramfort, Michel,
  Thirion, Grisel, Blondel, Prettenhofer, Weiss, Dubourg, Vanderplas, Passos,
  Cournapeau, Brucher, Perrot, \& Duchesnay}]{pedregosa2011}
Pedregosa, F., Varoquaux, G., Gramfort, A., {et~al.} 2011, Journal of Machine
  Learning Research, 12, 2825

\bibitem[{{Pilyugin} \& {Grebel}(2016)}]{pilyugin2016}
{Pilyugin}, L.~S. \& {Grebel}, E.~K. 2016, \mnras, 457, 3678

\bibitem[{{Pilyugin} \& {Mattsson}(2011)}]{pilyugin2011}
{Pilyugin}, L.~S. \& {Mattsson}, L. 2011, \mnras, 412, 1145

\bibitem[{{Pilyugin} {et~al.}(2010){Pilyugin}, {V{\'{\i}}lchez}, \&
  {Thuan}}]{pilyugin2010}
{Pilyugin}, L.~S., {V{\'{\i}}lchez}, J.~M., \& {Thuan}, T.~X. 2010, \apj, 720,
  1738

\bibitem[{{Piotrowska} {et~al.}(2022){Piotrowska}, {Bluck}, {Maiolino}, \&
  {Peng}}]{piotrowska2022}
{Piotrowska}, J.~M., {Bluck}, A. F.~L., {Maiolino}, R., \& {Peng}, Y. 2022,
  \mnras, 512, 1052

\bibitem[{{S{\'a}nchez} {et~al.}(2022){S{\'a}nchez}, {Barrera-Ballesteros},
  {Lacerda}, {Mej{\'\i}a-Narvaez}, {Camps-Fari{\~n}a}, {Bruzual},
  {Espinosa-Ponce}, {Rodr{\'\i}guez-Puebla}, {Calette}, {Ibarra-Medel},
  {Avila-Reese}, {Hernandez-Toledo}, {Bershady}, {Cano-Diaz}, \&
  {Munguia-Cordova}}]{sanchez2022}
{S{\'a}nchez}, S.~F., {Barrera-Ballesteros}, J.~K., {Lacerda}, E., {et~al.}
  2022, \apjs, 262, 36

\bibitem[{{S{\'a}nchez} {et~al.}(2019){S{\'a}nchez}, {Barrera-Ballesteros},
  {L{\'o}pez-Cob{\'a}}, {Brough}, {Bryant}, {Bland-Hawthorn}, {Croom}, {van de
  Sande}, {Cortese}, {Goodwin}, {Lawrence}, {L{\'o}pez-S{\'a}nchez}, {Sweet},
  {Owers}, {Richards}, \& {Walcher}}]{sanchez2019b}
{S{\'a}nchez}, S.~F., {Barrera-Ballesteros}, J.~K., {L{\'o}pez-Cob{\'a}}, C.,
  {et~al.} 2019, \mnras, 484, 3042

\bibitem[{{S{\'a}nchez} {et~al.}(2017){S{\'a}nchez}, {Barrera-Ballesteros},
  {S{\'a}nchez-Menguiano}, {Walcher}, {Marino}, {Galbany}, {Bland-Hawthorn},
  {Cano-D{\'{\i}}az}, {Garc{\'{\i}}a-Benito}, {L{\'o}pez-Cob{\'a}}, {Zibetti},
  {Vilchez}, {Igl{\'e}sias-P{\'a}ramo}, {Kehrig}, {L{\'o}pez S{\'a}nchez},
  {Duarte Puertas}, \& {Ziegler}}]{sanchez2017}
{S{\'a}nchez}, S.~F., {Barrera-Ballesteros}, J.~K., {S{\'a}nchez-Menguiano},
  L., {et~al.} 2017, \mnras, 469, 2121

\bibitem[{{S{\'a}nchez Almeida} \& {Dalla Vecchia}(2018)}]{sanchezalmeida2018}
{S{\'a}nchez Almeida}, J. \& {Dalla Vecchia}, C. 2018, \apj, 859, 109

\bibitem[{{S{\'a}nchez Almeida} {et~al.}(2014){S{\'a}nchez Almeida},
  {Elmegreen}, {Mu{\~n}oz-Tu{\~n}{\'o}n}, \& {Elmegreen}}]{sanchezalmeida2014}
{S{\'a}nchez Almeida}, J., {Elmegreen}, B.~G., {Mu{\~n}oz-Tu{\~n}{\'o}n}, C.,
  \& {Elmegreen}, D.~M. 2014, \aapr, 22, 71

\bibitem[{{S{\'a}nchez Almeida} \&
  {S{\'a}nchez-Menguiano}(2019)}]{sanchezalmeida2019}
{S{\'a}nchez Almeida}, J. \& {S{\'a}nchez-Menguiano}, L. 2019, \apjl, 878, L6

\bibitem[{{S{\'a}nchez-Menguiano} {et~al.}(2019){S{\'a}nchez-Menguiano},
  {S{\'a}nchez Almeida}, {Mu{\~n}oz-Tu{\~n}{\'o}n}, {S{\'a}nchez}, {Filho},
  {Hwang}, \& {Drory}}]{sanchezmenguiano2019}
{S{\'a}nchez-Menguiano}, L., {S{\'a}nchez Almeida}, J.,
  {Mu{\~n}oz-Tu{\~n}{\'o}n}, C., {et~al.} 2019, \apj, 882, 9

\bibitem[{{S{\'a}nchez-Menguiano} {et~al.}(2024){S{\'a}nchez-Menguiano},
  {S{\'a}nchez Almeida}, {S{\'a}nchez}, \&
  {Mu{\~n}oz-Tu{\~n}{\'o}n}}]{sanchezmenguiano2024}
{S{\'a}nchez-Menguiano}, L., {S{\'a}nchez Almeida}, J., {S{\'a}nchez}, S.~F.,
  \& {Mu{\~n}oz-Tu{\~n}{\'o}n}, C. 2024, \aap, 681, A121

\bibitem[{{Sanders} {et~al.}(2021){Sanders}, {Shapley}, {Jones}, {Reddy},
  {Kriek}, {Siana}, {Coil}, {Mobasher}, {Shivaei}, {Dav{\'e}}, {Azadi},
  {Price}, {Leung}, {Freeman}, {Fetherolf}, {de Groot}, {Zick}, \&
  {Barro}}]{sanders2021}
{Sanders}, R.~L., {Shapley}, A.~E., {Jones}, T., {et~al.} 2021, \apj, 914, 19

\bibitem[{{Scholte} \& {Saintonge}(2023)}]{scholte2023}
{Scholte}, D. \& {Saintonge}, A. 2023, \mnras, 518, 353

\bibitem[{{Tremonti} {et~al.}(2004){Tremonti}, {Heckman}, {Kauffmann},
  {Brinchmann}, {Charlot}, {White}, {Seibert}, {Peng}, {Schlegel}, {Uomoto},
  {Fukugita}, \& {Brinkmann}}]{tremonti2004}
{Tremonti}, C.~A., {Heckman}, T.~M., {Kauffmann}, G., {et~al.} 2004, \apj, 613,
  898

\bibitem[{{Wake} {et~al.}(2017){Wake}, {Bundy}, {Diamond-Stanic}, {Yan},
  {Blanton}, {Bershady}, {S{\'a}nchez-Gallego}, {Drory}, {Jones}, {Kauffmann},
  {Law}, {Li}, {MacDonald}, {Masters}, {Thomas}, {Tinker}, {Weijmans}, \&
  {Brownstein}}]{wake2017}
{Wake}, D.~A., {Bundy}, K., {Diamond-Stanic}, A.~M., {et~al.} 2017, \aj, 154,
  86

\bibitem[{{Wu} {et~al.}(2016){Wu}, {Zhang}, {Zhao}, \& {Zhang}}]{wu2016}
{Wu}, Y.-Z., {Zhang}, S.-N., {Zhao}, Y.-H., \& {Zhang}, W. 2016, \mnras, 457,
  2929

\bibitem[{{Yan} {et~al.}(2016){Yan}, {Bundy}, {Law}, {Bershady}, {Andrews},
  {Cherinka}, {Diamond-Stanic}, {Drory}, {MacDonald}, {S{\'a}nchez-Gallego},
  {Thomas}, {Wake}, {Weijmans}, {Westfall}, {Zhang}, {Arag{\'o}n-Salamanca},
  {Belfiore}, {Bizyaev}, {Blanc}, {Blanton}, {Brownstein}, {Cappellari},
  {D'Souza}, {Emsellem}, {Fu}, {Gaulme}, {Graham}, {Goddard}, {Gunn},
  {Harding}, {Jones}, {Kinemuchi}, {Li}, {Li}, {Maiolino}, {Mao}, {Maraston},
  {Masters}, {Merrifield}, {Oravetz}, {Pan}, {Parejko}, {Sanchez}, {Schlegel},
  {Simmons}, {Thanjavur}, {Tinker}, {Tremonti}, {van den Bosch}, \&
  {Zheng}}]{yan2016}
{Yan}, R., {Bundy}, K., {Law}, D.~R., {et~al.} 2016, \aj, 152, 197

\bibitem[{{Yates} {et~al.}(2020){Yates}, {Schady}, {Chen}, {Schweyer}, \&
  {Wiseman}}]{yates2020}
{Yates}, R.~M., {Schady}, P., {Chen}, T.~W., {Schweyer}, T., \& {Wiseman}, P.
  2020, \aap, 634, A107

\end{thebibliography}

\begin{appendix} 
\section{The effect of the adopted metallicity calibration}\label{sec:appendix1}
The determination of gas metallicity is a complex process fraught with numerous systematics and sources of uncertainty. Well-documented discrepancies exist when different strong-line calibrators are employed \citep[e.g.][]{kewley2008, lopezsanchez2012, kewley2019}. In this appendix we show that our results are not contingent upon the choice of the adopted gas-metallicity indicator. For that, we conducted a parallel analysis based on the oxygen abundance estimates from 15 alternative calibrations included in the {\tt pyPipe3D} VAC table\footnote{The O3O2, the R2, and the O3S2 calibrations proposed in \citet{curti2017} and also available in the {\tt pyPipe3D} VAC table were not used due to their limited dynamical range (see appendix B of Paper I for more details).}. All of these calibrators present a monotonic behaviour in the covered range of abundances of this study (8.2 < 12 + log(O/H) < 8.6), and therefore they allowed us to unambiguously measure the gas metallicity for all galaxies in the sample. Table~\ref{tabB1} lists the calibrations used and the corresponding references. 

\begin{table}[h]
\caption{List of alternative calibrations used to derive the oxygen abundances.}
\label{tabB1}      
\centering          
\begin{tabular}{c@{\hspace{1cm}}c}   
ID & Reference \\
\hline\\[-0.355cm]\hline\\[-0.3cm]
N2-M13 & \citet{marino2013}\\[0.1cm]
N2O2-KD02 & \citet{kewley2002}\\[0.1cm]
ONS-P10 & \citet{pilyugin2010}\\[0.1cm]
ON-P10 & \citet{pilyugin2010}\\[0.1cm]
NS-P11 & \citet{pilyugin2011}\\[0.1cm]
RS32-C17 & \citet{curti2017}\\[0.1cm]
R3-C17 & \citet{curti2017}\\[0.1cm]
S2-C17 & \citet{curti2017}\\[0.1cm]
N2-C17 & \citet{curti2017}\\[0.1cm]
R23-C17 & \citet{curti2017}\\[0.1cm]
O3N2-C17 & \citet{curti2017}\\[0.1cm]
R23-KK04 & \citet{kobulnicky2004}\\[0.1cm]
R-P16 & \citet{pilyugin2016}\\[0.1cm]
S-P16 & \citet{pilyugin2016}\\[0.1cm]
H19 & \citet{ho2019}\\[0.1cm]
\hline                  
\end{tabular}
\end{table}

\begin{figure}
\centering
\resizebox{\hsize}{!}{\includegraphics{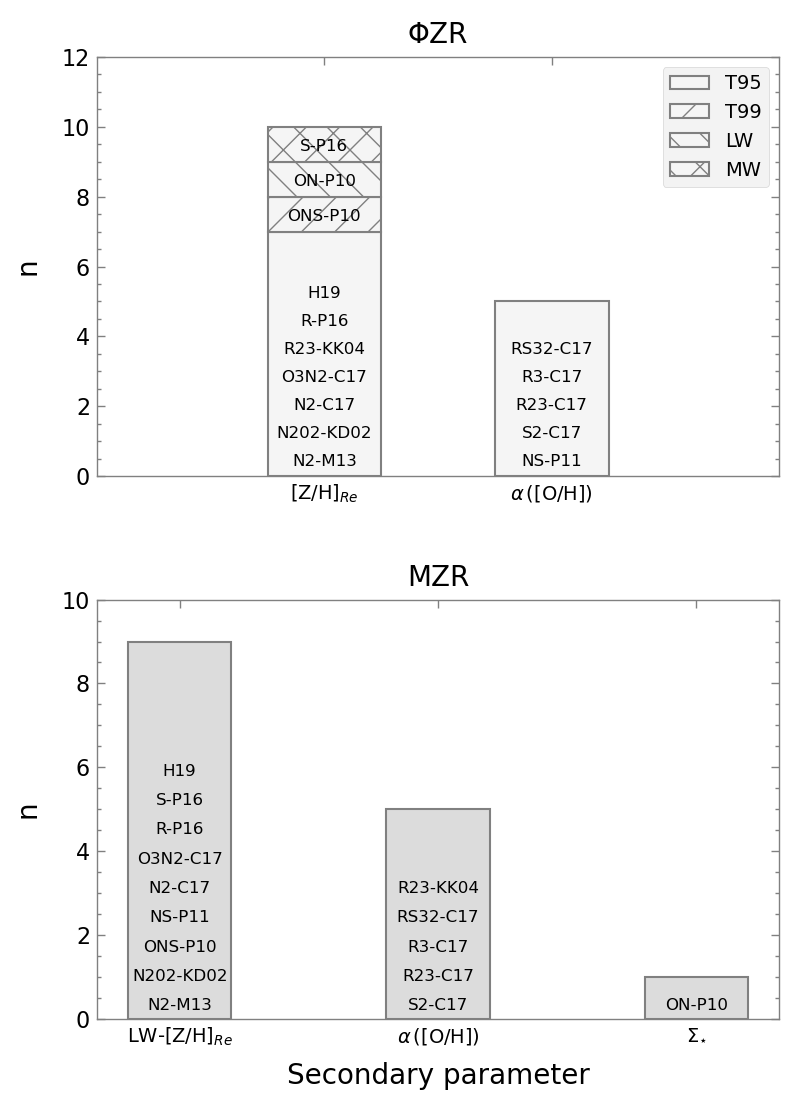}}
\caption{Histogram of the most relevant parameter in the RF when studying the residuals of the $\Phi$ZR (top panel) and the MZR (bottom panel) using 15 different estimators of $Z_g$ (see Table~\ref{tabB1} for references). For the top panel, LW (MW) means light- (mass-)weighted and T95 (T99) represents the look-back time at which the galaxy has formed 95\% (99\%) of its mass. All of these attributes refer to different estimations of the stellar metallicity measured at 1 $R_e$.}
\label{fig1apa}
\end{figure}

Similar to the procedure followed in Sec.~\ref{sec:results} for the main abundance calibrator, for each alternative indicator, we ran  a RF using the residual $Z_g$ ($\Delta Z_g$) from the $\Phi$ZR as the targets and removing $\Phi_{\rm baryon}$ from the input parameters. In the top panel of Figure~\ref{fig1apa}, we show a histogram with the most relevant parameter in the model according to its importance value in all 15 cases. We can see that in ten of them, an indicator of the global stellar metallicity is again the galactic property with the highest effect on $\Delta Z_g$. In the remaining five cases, a stronger dependence with the gas metallicity gradient ($\alpha$([O/H])) is found. We attribute this latter finding to the uncertainty in the determination of $R_e$, which can result in the $Z_g$ measured at this $R_e$ not being completely representative of the global gas metallicity (and thus the need to apply a small correction considering the existing negative gradient appears). Alternatively, we replicated this analysis with the residual $Z_g$ from the MZR, whose outcome is represented in a similar way in the bottom panel of Figure~\ref{fig1apa}. In this case, a secondary dependence of $Z_g$ with the global luminosity-weighted stellar metallicity is also reported for the majority of cases. We consider that the analyses described in this appendix reinforce the role of the stellar metallicity as the strongest secondary parameter shaping the gas metallicity in galaxies. 

\section{The $\Phi$ZR relation based on $M_\star / R_e^{\,0.6}$ as a tracer of the total gravitational potential}\label{sec:appendix2}
In Paper I we showed how the inclusion in the RF of a parameter of the type $M_\star / R_e^{\,\alpha}$ with $\alpha=0.6$ (different coefficients for $\alpha$ were explored) performs better than $\Phi_{\rm baryon} = M_\star / R_e$ when predicting $Z_g$. We investigated the effect of the dark matter (DM) halo on the gravitational potential and we argued how a scale of the form $M_\star / R_e^{\, 0.6}$ for the total $\Phi$ matches very well the theoretical relation between the DM fraction and the baryonic surface density predicted in cosmological numerical simulations of galaxy formation \citep{nestorshachar2023}. Thus it is important to confirm that the stellar metallicity plays the same role as the secondary parameter in the total $\Phi$ZR.

\begin{figure}
\centering
\resizebox{\hsize}{!}{\includegraphics{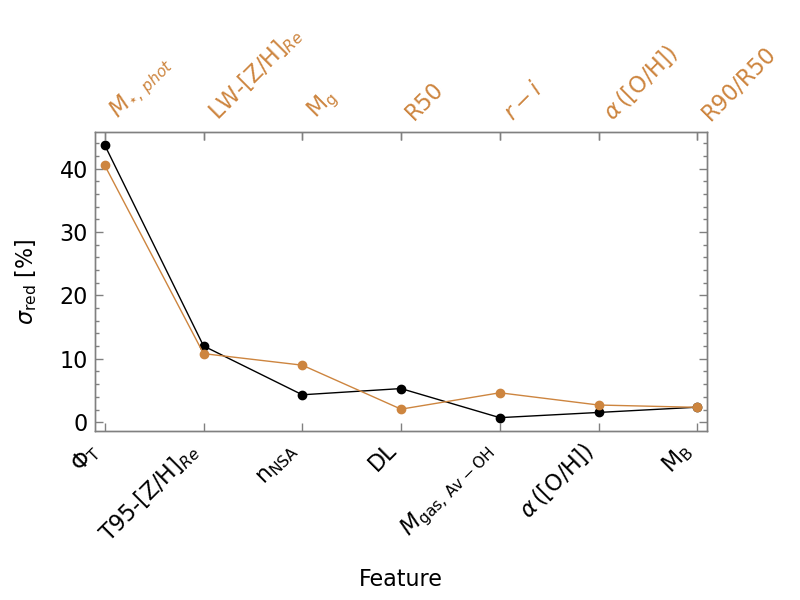}}
\caption{Analogous to Fig.~\ref{fig2} but including a tracer of the total gravitational potential in the model ($\Phi_{\rm T}$). See the caption of Fig.~\ref{fig2} for details.}
\label{fig1apb}
 \end{figure}

For this test, we included $M_\star / R_e^{\,0.6}$ (as a tracer of $\Phi_{\rm T}$) as an input parameter in the RF and we repeated the procedure followed in Sec.~\ref{sec:results}. The black line in Fig.~\ref{fig1apb} (bottom x-axis) represents the decrease in the dispersion of $Z_g$ as we modelled its relation with different galaxy parameters. Similar to Fig.~\ref{fig2}, the secondary x-axis (top) shows the case when we forced the stellar mass to be the first parameter to model (orange line). After fitting the dependence on $\Phi_{\rm T}$ ($44\%$), the RF again revealed the strong effect of the stellar metallicity (in particular, T95-[Z/H]$_{Re}$) on predicting $\Delta Z_g$, which is the parameter in the model with the highest importance value and the one that reduces the dispersion of the residual metallicities the most ($14\%$). The subsequent modelling of the residuals with the remaining galaxy properties produces a decrease in the dispersion always below $5\%$. This test confirms the main conclusion reached in our study, that is, the stellar metallicity is the most important secondary parameter in the $\Phi$ZR (both baryonic and total) and the MZR. 

\end{appendix}

\end{document}